\documentclass[prl,aps,twocolumn]{revtex4}

\usepackage{amsfonts}

\begin{document}

\newcommand{\capeq}[1]{Equation (\ref{eq:#1})}
\newcommand{\capsec}[1]{Section \ref{sec:#1}}
\newcommand{\eq}[1]{Eq.~(\ref{eq:#1})}
\renewcommand{\sec}[1]{Sec.~\ref{sec:#1}}
\newcommand\beq{\begin{equation}}
\newcommand\eeq{\end{equation}}
\newcommand\be{\begin{equation}}

\newcommand\ee{\end{equation}}
\newcommand\bea{\begin{eqnarray}}
\newcommand\eea{\end{eqnarray}}
\newcommand{\ket}[1]{| #1 \rangle}
\newcommand{\bra}[1]{\langle #1 |}
\newcommand{\braket}[2]{\langle #1 | #2 \rangle}
\newcommand{\proj}[1]{| #1\rangle\!\langle #1 |}
\newcommand{\ba}{\begin{array}}
\newcommand{\ea}{\end{array}}
\newtheorem{theorem}{Theorem}
\newtheorem{conjecture}{Conjecture}

\def\ms{\hspace*{-0.1ex}}
\def\ecall{{\cal E}}
\def\ecalun{{\cal E'}}
\def\rhol{\rho}
\def\rhoun{\rho'}
\def\lbm{ \left[\rule{0pt}{2.1ex}\right. }
\def\rbm{ \left.\rule{0pt}{2.1ex}\right] }
\def\lpm{ \left(\rule{0pt}{2.1ex}\right. }
\def\rpm{ \left.\rule{0pt}{2.1ex}\right) }
\def\lbL{ \left[\rule{0pt}{2.4ex}\right. }
\def\rbL{ \left.\rule{0pt}{2.4ex}\right] }
\def\lpL{ \left(\rule{0pt}{2.4ex}\right. }
\def\rpL{ \left.\rule{0pt}{2.4ex}\right) }
\def\ss{\hspace*{0.2ex}}
\def\<{\langle}
\def\>{\rangle}
\def\hcal{{\cal H}}
\def\bcal{{\cal B}}
\def\lcal{{\cal L}}
\def\ecal{{\cal E}}
\def\mcal{{\cal M}}
\def\iacc{I_{acc}}

\def\ot{\otimes}
\def\ra{\rightarrow}
\def\tr{{\rm Tr}}
\def\non{\nonumber}
\def\dag{\dagger}

\def\eqshort{\!=\!}
\def\rhorest{\rho_I}

\newtheorem{theo}{Theorem}
\newtheorem{obs}{Observation}
\newtheorem{defi}{Definition}
\newtheorem{lem}{Lemma}
\newtheorem{exam}{Example}
\newtheorem{prop}{Property}
\newtheorem{propo}{Proposition}
\newtheorem{cor}{Corollary}
\newtheorem{conj}{Conjecture}

\author{David P.~DiVincenzo$^{1,2}$, Micha{\l}
Horodecki$^{3}$, Debbie W.~Leung$^{1,2,4}$, John A.~Smolin$^1$,
and Barbara M.~Terhal$^{1,2}$\vspace*{1ex}}

\title{Locking classical correlation in quantum states}

\affiliation{$^1${IBM Watson Research Center, P.O. Box 218,
Yorktown Heights, New York 10598, USA}}

\affiliation{$^2${Institute for Quantum Information,
California Institute of Technology, Pasadena, California 91125-8100, USA}}

\affiliation{$^3${Institute of Theoretical Physics and
Astrophysics, Univ. of Gda\'nsk, 80--952 Gda\'nsk, Poland}}

\affiliation{$^4${Mathematical Science Research Institute, 1000
Centennial Driv'e, Berkeley, California 94720, USA} \vspace*{1ex}}
\date{\today}

\begin{abstract}
We show that there exist bipartite quantum states which contain
large hidden classical correlation that can be unlocked by a
disproportionately small amount of classical communication.  In
particular, there are $(2n+1)$-qubit states for which a one bit
message doubles the optimal classical mutual information between
measurement results on the subsystems, from $n/2$ bits to $n$
bits. States exhibiting this behavior need not be entangled.  We
study the range of states exhibiting this phenomenon and bound its
magnitude.
\end{abstract}


\maketitle

\raggedbottom

The study of possible correlations between quantum systems
was initiated by Einstein, Podolsky and Rosen \cite{EPR} and
Schr\"odinger \cite{Schroedinger}.
These pioneers were concerned with entanglement --- quantum
correlation that are non-existent in classical physics.
Recent development in quantum information theory has motivated
extensive study of entanglement (see \cite{QIC-review} for a review).
Furthermore, an exciting subject of characterizing other interesting
types of correlations has emerged.
For example, quantum correlation, classical one, or quantum {\it
and} classical correlation have been studied
\cite{Zurek-Annalen,HendersonVedral,Zurek-discord-01,OHHH2001,terhal+:epur}.

The classical mutual information of a quantum state $\rho_{AB}$
can be defined naturally ~\cite{terhal+:epur} as the maximum
classical mutual information that can be obtained by local
measurements $M_A\ot M_B$ on the state $\rho_{AB}$:
\be
    I_c(\rho) \equiv \max_{M_A\ot M_B} I(A \!:\! B).
\label{eq:icdef}
\ee
Here $I(A \!:\! B)$ is the classical mutual information defined as
$I(A\!:\!\!B) \equiv H(p_A) + H(p_B) - H(p_{AB})$, $H$ is the
entropy function \cite{cover&thomas:infoth}, and $p_{AB},p_A,p_B$ are
the probability distributions of the joint and individual outcomes of
performing the local measurement $M_A\!  \ot\! M_B$ on $\rho$.
The physical relevance of $I_c$ is many-fold.
First, $I_c(\rho)$ is the maximum classical correlation obtainable
from $\rho$ by purely local processing. Second, $I_c(\rho)$
corresponds to the classical definition when $\rho$ is
``classical,'' i.e., diagonal in some local product basis and
corresponds to a classical distribution.
Third, when $\rho$ is pure, $I_c(\rho)$ is the correlation defined
by the Schmidt basis and thus equal to the entanglement of the
pure state~\cite{Peres93a,proof2}.
Finally $I_c(\rho)=0$ if and only if $\rho = \rho_A \ot
\rho_B$~\cite{proof5}.

Any good correlation measure should satisfy certain axiomatic
properties.
First, correlation is a nonlocal property and should not increase
under local processing ({\it monotonicity}) (I).
Second, a protocol starting from an uncorrelated initial state and
using $l$ qubits or $2l$ classical bits of communication (one-way
or two-way) and local operations should not create more than $2l$
bits of correlation.  We call this property {\it total
proportionality} (II). The intuition is that if $2l$ bits of
correlation can be established with fewer than $2l$ bits of
communication, then it may be possible to establish nonzero
correlation with no communication if the receiver guesses the
message.

We may expect other properties for any correlation measure.
If a protocol has several rounds of communication, one may consider
the increase of correlation due to each round of communication.
Intuitively, a small amount of communication should not increase
correlation abruptly.
In particular, one may expect that the transmission of $l$ qubits
or $2l$ bits should not increase the correlation of {\em any
initial state} by more than $2l$ bits. We call this property {\it
incremental proportionality} (III).
This strengthens total proportionality by allowing all possible
initial states, or equivalently by considering the increase in
correlation step-wise.  Other properties such as continuity in $\rho$
are also expected (IV).

All of these properties (I-IV) hold for some well known correlation
measures. They hold for the classical mutual information $I(A\!:\!B)$
when communication is classical~\cite{proof4} as one may expect.  They
also hold for the quantum mutual information $I_q(\rho)$
\cite{terhal+:epur} (for any communication).  Here $I_q(\rho)\equiv
S(\rho_A)\!+\!S(\rho_B)\!-\!S(\rho)$ with $S(\rho)\equiv -\tr \rho
\log \rho$ being the von Neumann entropy and $\rho_A=\tr_{\!B} \,
\rho$, $\rho_B=\tr_{\! A} \,\rho$. In Ref. \cite{terhal+:epur}
monotonicity, total proportionality, and continuity have been proved
for $I_c$, while incremental proportionality was only proved for
pure initial state $\rho$ (for any communication) and for the
classical restriction.

In this paper, we report the surprising fact that {\it incremental
proportionality} for $I_c$ can be violated in some extreme manner
for a mixed initial state $\rho$.  We will see that a single
classical bit, sent from Alice to Bob, can result in an {\em
arbitrarily large} increase in $I_c$.  This phenomenon can be
viewed as a way of locking classical correlation in the quantum
state $\rho$.
If one-bit of communication increases $I_c$ by a large amount, the
correlation must be ``present'' initially, though hidden or locked as
indicated by a small initial value of $I_c$.  Only after the one-bit
transmission can the large amount of correlation become accessible or
unlocked.
Since incremental proportionality of $I_c$ holds classically, the
phenomenon of locked correlation is a purely quantum effect.  It
is a direct consequence of the indistinguishability of
non-orthogonal quantum states.  Applications of such
indistinguishability are well known, most notably in quantum key
distribution \cite{BB84} and the various partial quantum bit
commitment and coin tossing protocols (see \cite{SR:bc,SR:qt} and
references therein). Curiously, the simple effect that we observe
and bound in this paper had not been noted before.

For a given initial state $\rho$ and the amount and type of
communication, we can capture the increase in correlation by
defining the following functions: \be
    I_c^{(l)}(\rho) = \max_{\Lambda^{(l)}} I_c(\Lambda^{(l)}(\rho))\,,~
    I_c^{[l]}(\rho) = \max_{\Lambda^{[l]}} I_c(\Lambda^{[l]}(\rho))\,.
\label{eq:1way}
\ee
The operator $\Lambda$ denotes a bipartite quantum operation consists
of local operations and no more than $l$ bits or qubits of
communication, a constraint denoted by the superscript $(l)$ or $[l]$.
Note that $I_c(\rho)=I_c^{(0)}(\rho)=I_c^{[0]}(\rho)$.
Throughout the paper, we use $\rhol$ and $\rhoun$ to denote the states
before and after the quantum operation with communication,
$\rhoun=\Lambda(\rhol)$.

With this notation, we summarize our main results:

\noindent $\bullet$ We present an example in which $1$ bit of
classical communication increases $I_c$ by ${1\over 2}\log d$ bits,
where $\rho$ consists of $1+\log d$ and $\log d$ qubits in Alice and
Bob's systems respectively.  Since $I_c$ satisfies total
proportionality, the classical correlation can be viewed as being locked
in the state $\rho$ and then unlocked in $\rho'$ by the $1$-bit
message.

\noindent $\bullet$ We bound the extent of incremental proportionality
violation in terms of the amount of initial correlation and the amount
of communication.  The amount of correlation unlocked by $l$ bits of
1-way classical communication can be bounded as (Theorem 1)
\be
    I_c^{(l)}(\rho) - I_c(\rho) ~ \leq ~
            l + (2^l \! - \!1) \, I_c(\rho) \,.
\label{theeq}
\ee
For small $I_c(\rho)$, the amount unlocked by $l$ qubits (two-way) can
be bounded as (Theorem 2)
\bea
    I_c^{[l]}(\rho) - I_c(\rho) ~ \leq ~ 2l + O(d^2 \sqrt{I_c(\rho)}\log
    I_c(\rho)) \,.
\label{eq:tomo}
\eea

We now describe the example in which an arbitrary amount of
correlation is unlocked with a one-bit message.
The initial state $\rhol$ is shared between subsystems held by Alice
and Bob, with respective dimensions $2d$ and $d$,
\be
    \rhol = {1 \over 2d}
    \sum_{k=0}^{d-1} \sum_{t=0}^1 (|k\>\<k| \ot |t\>\<t|)_A
    \ot (U_t |k\>\<k| U_t^\dag)_B
\,.
\label{thestate}
\ee
Here $U_0 = I$ and $U_1$ changes the computational basis to a
conjugate basis ($\forall_{i,k} \; |\bra{i} U_1
\ket{k}|=\frac{1}{\sqrt{d}}$).
In this example, Bob is given a random draw $|k\>$ from $d$ states in
two possible random bases (depending on $t=0$ or $1$), while Alice has
complete knowledge of his state.
To achieve $I_c^{(1)}(\rhol)=\log d + 1$, Alice sends $t$ to Bob, who
then undoes $U_t$ on his state and measures $k$ in the computational
basis.  Alice and Bob now share both $k$ and $t$, with $\log d + 1$
bits of correlation.

For example, the state $\rho$ can arise from the following scenario.
Let $d=2^n$.  Alice picks a random $n$-bit string $k$ and sends Bob
$\ket{k}$ or $H^{\otimes n} \ket{k}$ depending on whether the random
bit $t=0$ or $1$.  Here $H$ is the Hadamard transform.  Alice can send
$t$ to Bob to unlock the correlation later.
Experimentally, Hadamard transform and measurement on single qubits
are sufficient to prepare the state $\rho$ and later extract the
unlocked correlation in $\rhoun$ -- they can be realized using photons
and linear optical elements like quarter-wave plates and calcite
crystals.

Now we prove that $I_c(\rhol) = {1 \over 2} \log d$.
First, the complete measurement $M_{A}$ along $\{|k\> \ot
|t\>\}$ is provably optimal for Alice:  Since the outcome tells her
precisely which pure state from the ensemble she has, she
can apply {\em classical, local} post-processing to obtain the output
distribution for any other measurement she could have performed.
For Alice's choice of optimal measurement, $I_c(\rhol)$ is simply
Bob's {\em accessible} information $I_{\rm acc}$~\cite{Peres93a}
about the uniform ensemble of states
$\{|k\>,U_1|k\>\}_{k=0,\cdots,d-1}$.

In general, the accessible information $I_{\rm acc}$ about an ensemble
of states ${\cal E} = \{p_i \geq 0,\eta_i \}$ is the maximum mutual
information between $i$ and the outcome of a measurement.
$I_{\rm acc}({\cal E})$ can be maximized by a POVM with rank $1$
elements only~\cite{Peres93a}.
Let $M = \{\alpha_j |\phi_j\>\<\phi_j|\}_j$ stand for a POVM with rank
$1$ elements where each $|\phi_j\>$ is normalized and $\alpha_j >0$.
Then $I_{\rm acc}({\cal E})$ can be expressed as
\bea
    I_{\rm acc}({\cal E}) & = & \max_M \lbL
    -\sum_i p_i \log p_i
\label{eq:acc}
\\
\non
    & + & \sum_i \sum_j p_i \alpha_j \<\phi_j|\eta_i|\phi_j\>
        \, \log {p_i \<\phi_j|\eta_i|\phi_j\>
        \over \<\phi_j | \mu | \phi_j\>} \rbL
\,,
\eea
where $\mu = \sum_i p_i \eta_i$.

We now apply \eq{acc} to the present problem.  Our ensemble is $\{{1
\over 2d}, U_t |k\>\}_{k,t}$ with $i=k,t$, $p_{k,t} = {1 \over 2d}$, $\mu
= {I \over d}$, and $\<\phi_j | \mu | \phi_j\> = {1 \over d}$.
Putting all these in \eq{acc},
\bea
    I_c(\rhol) = \max_M \! \lbL \! \log 2d
    + \! \sum_{jkt} {\alpha_j \over 2d}
    |\<\phi_j|U_t |k\>|^2
    \log {|\<\phi_j|U_t|k\>|^2 \over 2} \! \rbL
\non
\\
    = \max_M \! \lbL \! \log d
    \! + \! \sum_j \! {\alpha_j \over d} \!
    \lpL \!\!{1 \over 2} \! \sum_{kt}  \!
    |\<\phi_j|U_t|k\>|^2
    \log |\<\phi_j|U_t|k\>|^2 \!\! \rpL \!\! \rbL
\non
\eea
where we use $\sum_j \alpha_j = d$ and $\forall_{jt} \sum_{k}
|\<\phi_j|U_t|k\>|^2 = 1$ to obtain the last line.
Since $\sum_j {\alpha_j \over d} = 1$, the second term is a convex
combination, and can be upper bounded by maximization over just one
term:
\bea
    \!\!\! I_c(\rhol) \leq \log d +
    \max_{|\phi\>} {1 \over 2} \sum_{kt}
    |\<\phi|U_t|k\>|^2 \log| \<\phi|U_t|k\>|^2
\ss .
\label{eq:iaccgen}
\eea
Note that $-\sum_{kt} |\<\phi|U_t|k\>|^2 \log| \<\phi|U_t|k\>|^2$ is
the sum of the entropies of measuring $|\phi\>$ in the computational
basis and the conjugate basis.
Reference \cite{MU:genent} proves that such a sum of entropies is at
least $\log d$.
Lower bounds of these type are called entropic uncertainty
inequalities, which quantify how much a vector $|\phi\>$ cannot be
simultaneously aligned with states from two conjugated bases.
It follows that $I_c(\rhol) \leq {1 \over 2} \log d$.
Equality can in fact be attained when Bob measures in the
computational basis, so that $I_c(\rhol) = {1 \over 2} \log d$ and
$I_c^{(1)}(\rhol) - I_c(\rhol) = 1 + {1 \over 2} \log d$.

We remark that incremental proportionality remains violated for
multiple copies of $\rho$.  Wootters proved that \cite{DLT:hiding}
the accessible information from $m$ independent draws of an
ensemble ${\cal E}$ of separable states is additive,
$I_{acc}({\cal E}^{\otimes m})=mI_{acc}({\cal E})$.  It follows
$I_c(\rhol^{\otimes m})=mI_c(\rhol)$ in our example.

One would expect a stronger locking effect when the message (a key) is
longer than one bit.  There are two figures of merit:
First, the ``amplification'' of correlation,  
$r_1 = I_c(\rhoun)/I_c(\rhol)$, should be large.
Second, the amount of unlocked information, compared to the key size,
$r_2=(I_c(\rhoun) - I_c(\rhol))/l$, should be large.
Ideally, we want both $r_1$ and $r_2$ to be arbitrarily large.
We have investigated (see the Appendix for details) this
possibility by generalizing our $2$-bases example to $L>2 $
conjugate (or mutually unbiased) bases.  The key size is then $l =
\log L$. We have found rigorous results for the two extreme cases,
namely the previous example with $L=2$ in which $(r_1,r_2) \approx
(2,\log d)$ and the case of $L=d\!+\!1$ bases
in which $(r_1,r_2) \approx (2 \log d,2)$.  We
believe some intermediate values of $L$ will make both $r_1,r_2$
large. For example, any $\log L=o(\log d)$ will guarantee that
$r_1$ is large. But an analytic proof that $r_2$ is also large has
proved to be difficult, and numerical studies are inconclusive
(see Appendix).

An even stronger kind of locking would be what we call {\em complete}
locking, in which $I_c(\rhol)$ would decrease rapidly with the key
size $l$, yet the key can retrieve a finite fraction of the
data. For example,
\bea
    I_c(\rhol) \propto 2^{-\alpha l}  ~~\mbox{and}~~
    I_c(\rhoun)-l \approx \delta \log d \,.
\label{eq:complete2}
\eea
where $\rho$ is supported on two $d$-dimensional systems, $\delta > 0$
is independent of $d,l$, and $\alpha > 0$.  Note that $r_1$, $r_2$ are
automatically large for large $d$ in complete locking.
We find that for large $d$ complete locking cannot occur with
$\alpha \geq 1$ or for very short keys $l=o( \log \log d)$.  This
follows from the following Theorem:

{\it Theorem 1}~~
If $\rho'$ is obtained from $\rho$ with $l$ bits of 1-way classical
communication, $I_c(\rhol) \geq 2^{-l}(I_c(\rhoun)-l)$.  It follows
$I_c^{(l)}(\rhol) - I_c(\rhol) \leq l + (2^l-1) I_c(\rhol)$.
%
\\[0.8ex]
The intuition behind the proof is that Bob can just guess the
classical key.
If he guesses correctly (with probability ${1\over 2^l}$), he gains
$I_c(\rhoun)$ bits of information, so that the average information
gain is at least ${1\over 2^l} I_c(\rhoun)$.

\vspace*{0.7ex}

\noindent {\em Proof}:
Let $\rhoun$ results from sending an $l$-bit message (or key) from
Alice to Bob.
Let the random variable $\tilde{Z}$ describe the key, and the random
variable $X$ be the outcome of Alice's POVM measurement that optimizes
$I_c(\rho')$.
We can always include $\tilde{Z}$ as part of $X$.
Bob applies one of $2^l$ possible measurements based on a random
variable $Z$, yielding the outcome $Y$.
To achieve $I_c(\rho')$, Bob takes $Z = \tilde{Z}$, and each of his
measurements is optimal for each value of $\tilde{Z}$.
Therefore~\cite{footshort}:
\be
\label{icrhoun}
    I_c(\rhoun)=I(X \!:\!Y \!\tilde{Z}Z|Z={\tilde Z})=
    I(X \!:\!Y \!\tilde{Z}|Z={\tilde Z})
\,.
\ee
Applying the chain rule~\cite{cover&thomas:infoth}:
\bea
\nonumber
      I(X \!:\! Y\!\tilde{Z}\,|\,Z\eqshort\tilde{Z})
    = I(X \!:\! Y|\,\tilde{Z},Z\eqshort\tilde{Z})
    + I(X \!:\!\tilde{Z}|Z\eqshort\tilde{Z})
\\
\label{ixyzzeqz}
    \le I(X\!:\!Y|\,\tilde{Z},Z\eqshort\tilde{Z})+l \hspace*{14ex}
\eea
where we have used $I(X \!:\! \tilde{Z}|Z\eqshort \tilde{Z})\le l$
because $l$ is the size of the key $\tilde{Z}$.

Working from the other end, consider the following not necessarily
optimal measurement on $\rho$:
Alice's measurement is same as before, but $\tilde{Z}$ is not sent
to Bob.
Instead, Bob draws $Z$ at random.  The resulting mutual
information provides a lower bound on $I_c(\rhol)$,
\mbox{$I_c(\rhol) \ge I(X\!:\!Y\!Z)$.}  By the chain rule, we can
write \mbox{$I(X\!:\!Y\!Z)=I(X\!:\!Y|Z)+I(X\!:\!Z) =
I(X\!:\!Y|Z)$}. Because $Z$ is independent of $X$ we have
\be
\label{ixybarz}
    I_c(\rhol) \ge I(X \!: \! Y|Z) \,.
\ee
Because $\tilde{Z}$ is part of $X$, we can write
\bea
\nonumber
    I(X\!:\!Y|Z) = I(X\tilde{Z}\!:\!Y|Z) \hspace*{20ex}
\\
    = I(X\!:\!Y|\tilde{Z}Z) + I(\tilde{Z}\!:\!Y|Z)
    \ge I(X\!:\!Y|\tilde{Z}Z)
\,,
\label{ixyzz}
\eea
again using the chain rule and \mbox{$I(\tilde{Z}\!:\!Y|Z)\ge 0$.}

Now, comparing (\ref{ixyzzeqz}) and (\ref{ixyzz}), 
\bea
\label{sum2}
    \hspace*{-2ex}
    I(X\!:\!Y|\,\tilde{Z},Z\eqshort\tilde{Z}) \! = \!\! \sum_{z_0} \!
    \Pr(\tilde{Z}\eqshort z_0) \,
    I(X\!\!:\!Y|\tilde{Z}\eqshort Z\eqshort z_0)
\ss ,
\\
\label{sum1}
    \hspace*{-2ex}
    I(X\!:\!Y| \, \tilde{Z} \! Z)
    = \! \! \sum_{z_0,z_1} \!\! \frac{\Pr(\tilde{Z}\eqshort z_0)
    I(X\!:\!Y|\tilde{Z}\eqshort z_0,Z\eqshort z_1)}{2^l} 
\,.~
\eea
The sum (\ref{sum1}) is the same sum as (\ref{sum2}) but with some
extra terms and a factor of $1/2^l$, so
\be
    I(X\!:\!Y|\tilde{Z}Z)
    \ge \frac{1}{2^l} I(X\!:\!Y|\tilde{Z},Z\eqshort \tilde{Z})
\,,
\label{penultimate}
\ee
and putting together (\ref{penultimate}) and
(\ref{icrhoun},\ref{ixyzzeqz},\ref{ixybarz}) proves the first statement.
The second statement is true because there is only one round of
communication; monotonicity then implies the optimal $\Lambda^{(l)}$
in (\ref{theeq}) consists of just the communication.  $\Box$


We can bound the violation of incremental proportionality in yet
another way.  Total proportionality for $I_c$ (when $I_c(\rho) = 0$,
transmitting $l$ qubits can increase $I_c$ by at most $l$ bits) can be
restated as ``$I_c(\rho) = 0$'' implies no incremental proportionality
violation.  We may thus expect a small violation of incremental
proportionality when $I_c(\rho)$ is small.  We are able to prove the
following: \\[1.2ex]
{\it Theorem 2}~~
Let $\rho$ be a bipartite state on $C^d \otimes C^d$ and $\rho'$ be
obtained from $\rho$ by $l$ qubits of two-way communication.  If
$I_c(\rho) \leq {1\over 6\ln 2}{1\over (d+1)^2}$,
\bea I_c(\rho') -  I_c(\rho) \hspace*{38ex}
\non
\\
\leq 2l - (2d)^2 \!\!
\sqrt{\ms (2 \ln 2) I_{\ms c \ms}(\ms\rho) } \,
\log \! \sqrt{\ms (2 \ln 2) I_{\ms c \ms}(\ms\rho)} \, .
\non
\eea

\noindent The proof of theorem 2 relies essentially on the
following lemma (see the Appendix for a proof) which says that
when $I_c(\rho)$ is small, $\rho$ must be close to an uncorrelated
state (in trace
distance). \\[1.2ex]
{\it Lemma 1}~~
If $\rho$ is a bipartite state on $C^{d} \otimes
C^d$, then
\bea
    \hspace*{2ex} \tr \ss | \ss \rho_{AB} - \rho_A \ot \rho_B|
    \leq (2d)^2 \sqrt{2 \ln 2 ~ I_c(\rho)}
\label{eq:trace-Ic-main} \,,
\eea
where $\rho_{A/B}={\rm Tr}_{B/A}\rho$. \\[1.2ex]
The theorem can be proved by first relating $I_c$ to $I_q$ which obeys
incremental proportionality (with an extra factor of $2$).  Then Lemma
1 and the continuity of $I_q$ implies $I_q(\rho)$ is
close to $I_q(\rho_A \otimes \rho_B)$, giving the desired bound (see
the Appendix for details).

The weakness of Lemma 1 and thus that of Theorem 2 stems from the
factor $d^2$ in Lemma 1.
This factor comes from an analysis that uses measurements in all
mutually unbiased bases to distinguish $\rho_A \otimes \rho_B$
from $\rho$, and the analysis is probably not optimal. Note that
the dependence on the dimension $d$ in the bound in Theorem 2
makes it impossible to completely rule out complete locking.

Our locking scheme is closely related to quantum key distribution
(QKD), in particular BB84 \cite{BB84}, in which Alice holds a basis bit
(computational or Hadamard) for each of Bob's qubits.
Transmitting the locked state limits the classical correlation between
Alice and any potential eavesdropper (Eve) and forbids her from
tampering without disturbance.
Announcing the basis bits at a later stage enables Alice and Bob to
unlock the correlation.  Furthermore, incomplete unlocked correlation
(as indicated by the test bits) reveals Eve's tampering.
However, in BB84, one bit is sent for every bit to be unlocked, and
there is no extreme unlocking behavior as shown by our examples.

Further research into the phenomenon of locking will be worthwhile.
For instance, we have seen differences in the locking effect by
quantum and classical keys.
Another important factor affecting the strength of locking is the
number of rounds of communication allowed.
In fact, a striking difference between one-way and two-way
communications can be seen if one generalize the state in Eq.
(\ref{thestate}) so each of Alice and Bob has a one-bit key register,
and the rotation $U_t$, now performed on both Bob's and Alice's state,
is determined by the {\em parity} of the two key bits.  Full unlocking
is possible with two-way communication, but not with one-way
communication.
Finally, the possibility of complete locking, or the impossibility
(by improving Lemma 1 and Theorem 2) are important open
questions; it may be interesting to see how complete locking relates
to known restrictions on partial bit commitments \cite{SR:bc}.

{\bf Acknowledgments}: We thank I. Devetak and C. Bennett for
extremely helpful discussions, W. Wootters for enlightening discussion
on mutually unbiased bases, and D. Gottesman for a discussion on
complete locking.  Part of this work was completed while MH was
visiting at the MSRI program on Quantum Computation.  MH is supported
by EC, contract No.~IST-2001-37559 (RESQ) and grant QUPRODIS, and also
by IST-1999-11053 (EQUIP).
DPDV, JAS and BMT are supported in part by the NSA and the ARDA
through ARO contract~No. DAAD19-01-C-0056.  DWL acknowledges support
by a Tolman fellowship and support from the NSF under Grant
No.~EIA-0086038.


\begin{thebibliography}{25}
\expandafter\ifx\csname natexlab\endcsname\relax\def\natexlab#1{#1}\fi
\expandafter\ifx\csname bibnamefont\endcsname\relax
  \def\bibnamefont#1{#1}\fi
\expandafter\ifx\csname bibfnamefont\endcsname\relax
  \def\bibfnamefont#1{#1}\fi
\expandafter\ifx\csname citenamefont\endcsname\relax
  \def\citenamefont#1{#1}\fi
\expandafter\ifx\csname url\endcsname\relax
  \def\url#1{\texttt{#1}}\fi
\expandafter\ifx\csname urlprefix\endcsname\relax\def\urlprefix{URL }\fi
\providecommand{\bibinfo}[2]{#2}
\providecommand{\eprint}[2][]{\url{#2}}

\bibitem[{\citenamefont{Einstein et~al.}(1935)\citenamefont{Einstein,
Podolsky, and Rosen}}]{EPR}
\bibinfo{author}{\bibfnamefont{A.}~\bibnamefont{Einstein}},
\bibinfo{author}{\bibfnamefont{B.}~\bibnamefont{Podolsky}},
\bibnamefont{and}
\bibinfo{author}{\bibfnamefont{N.}~\bibnamefont{Rosen}},
\bibinfo{journal}{Phys. Rev.} \textbf{\bibinfo{volume}{47}},
\bibinfo{pages}{777} (\bibinfo{year}{1935}).

\bibitem[{\citenamefont{Schr\"odinger}(1935)}]{Schroedinger}
\bibinfo{author}{\bibfnamefont{E.}~\bibnamefont{Schr\"odinger}},
\bibinfo{journal}{Naturwissenschaften} \textbf{\bibinfo{volume}{23}},
\bibinfo{pages}{807} (\bibinfo{year}{1935}).

\bibitem[{QIC()}]{QIC-review}
\bibinfo{note}{Special issue of Q. Info. Comp., vol. 1, 2001}.

\bibitem[{\citenamefont{Zurek}(2000)}]{Zurek-Annalen}
\bibinfo{author}{\bibfnamefont{W.}~\bibnamefont{Zurek}},
\bibinfo{journal}{Ann.  Phys.} \textbf{\bibinfo{volume}{9}},
\bibinfo{pages}{5} (\bibinfo{year}{2000}).

\bibitem[{\citenamefont{Henderson and Vedral}()}]{HendersonVedral}
\bibinfo{author}{\bibfnamefont{L.}~\bibnamefont{Henderson}}
\bibnamefont{and}
\bibinfo{author}{\bibfnamefont{V.}~\bibnamefont{Vedral}},
\eprint{quant-ph/0105028}.

\bibitem[{\citenamefont{Ollivier and Zurek}(2002)}]{Zurek-discord-01}
\bibinfo{author}{\bibfnamefont{H.}~\bibnamefont{Ollivier}}
\bibnamefont{and}
\bibinfo{author}{\bibfnamefont{W.}~\bibnamefont{Zurek}},
\bibinfo{journal}{Phys. Rev. Lett.} \textbf{\bibinfo{volume}{88}},
\bibinfo{pages}{17901} (\bibinfo{year}{2002}),
\eprint{quant-ph/0105072}.

\bibitem[{\citenamefont{Oppenheim
et~al.}(2002)\citenamefont{Oppenheim, Horodecki, Horodecki, and
Horodecki}}]{OHHH2001}
\bibinfo{author}{\bibfnamefont{J.}~\bibnamefont{Oppenheim}},
\bibinfo{author}{\bibfnamefont{M.}~\bibnamefont{Horodecki}},
\bibinfo{author}{\bibfnamefont{P.}~\bibnamefont{Horodecki}},
\bibnamefont{and}
\bibinfo{author}{\bibfnamefont{R.}~\bibnamefont{Horodecki}},
\bibinfo{journal}{Phys. Rev. Lett.} \textbf{\bibinfo{volume}{89}},
\bibinfo{pages}{180402} (\bibinfo{year}{2002}),
\eprint{quant-ph/0112074}.

\bibitem[{\citenamefont{Terhal et~al.}(2002)\citenamefont{Terhal,
Horodecki, Leung, and DiVincenzo}}]{terhal+:epur}
\bibinfo{author}{\bibfnamefont{B.}~\bibnamefont{Terhal}},
\bibinfo{author}{\bibfnamefont{M.}~\bibnamefont{Horodecki}},
\bibinfo{author}{\bibfnamefont{D.}~\bibnamefont{Leung}},
\bibnamefont{and}
\bibinfo{author}{\bibfnamefont{D.}~\bibnamefont{DiVincenzo}},
\bibinfo{journal}{J. Math. Phys.} \textbf{\bibinfo{volume}{43}},
\bibinfo{pages}{4286} (\bibinfo{year}{2002}),
\bibinfo{note}{quant-ph/0202044}.

\bibitem[{\citenamefont{Cover and Thomas}(1991)}]{cover&thomas:infoth}
\bibinfo{author}{\bibfnamefont{T.}~\bibnamefont{Cover}}
\bibnamefont{and}
\bibinfo{author}{\bibfnamefont{J.}~\bibnamefont{Thomas}},
\emph{\bibinfo{title}{Elements of Information Theory}}
(\bibinfo{publisher}{Wiley, New York}, \bibinfo{year}{1991}).

\bibitem[{\citenamefont{Peres}(1993)}]{Peres93a}
\bibinfo{author}{\bibfnamefont{A.}~\bibnamefont{Peres}},
\emph{\bibinfo{title}{Quantum Theory: Concepts and Methods}}
(\bibinfo{publisher}{Kluwer Academic}, \bibinfo{address}{Dordrecht},
\bibinfo{year}{1993}).

\bibitem[{pro({\natexlab{a}})}]{proof2} \bibinfo{note}{When $\rho$ is
classical (pure), the measurement along the local product basis
(Schmidt basis) is optimal by the data processing
inequality~\cite{cover&thomas:infoth} and the fact other measurement
outcomes are obtainable by local processing of the optimal one.}

\bibitem[{pro({\natexlab{b}})}]{proof5} \bibinfo{note}{Obviously,
$I_c(\rho_A \ot \rho_B) = 0$. The converse follows from
Eq.~(\ref{eq:trace-Ic-main}) in Lemma 1.}

\bibitem[{pro({\natexlab{c}})}]{proof4} \bibinfo{note}{For example,
(total and) incremental proportionality of $I(A:B)$ for the classical
case follows from the fact~\cite{cover&thomas:infoth} that
$\max(H(p_{A}),H(p_{B})) \leq H(p_{AB}) \leq H(p_{A})+H(p_{B})$, so
that when Alice sends a classical system $A'$ to Bob, $I_c(\rho') =
I(A;BA') \leq I(AA';B)+H(p_{A'})$.}

\bibitem[{\citenamefont{Bennett and Brassard}(1984)}]{BB84}
\bibinfo{author}{\bibfnamefont{C.}~\bibnamefont{Bennett}}
\bibnamefont{and}
\bibinfo{author}{\bibfnamefont{G.}~\bibnamefont{Brassard}},
\bibinfo{journal}{Proc. of IEEE Int. Conference on Computers, Systems
and Signal Processing, Bangalore, India (IEEE, New York)}
p. \bibinfo{pages}{175} (\bibinfo{year}{1984}).

\bibitem[{\citenamefont{Spekkens and Rudolph}(2001)}]{SR:bc}
\bibinfo{author}{\bibfnamefont{R.}~\bibnamefont{Spekkens}}
\bibnamefont{and}
\bibinfo{author}{\bibfnamefont{T.}~\bibnamefont{Rudolph}}
(\bibinfo{year}{2001}), \bibinfo{note}{quant-ph/0106019}.

\bibitem[{\citenamefont{Spekkens and Rudolph}(2002)}]{SR:qt}
\bibinfo{author}{\bibfnamefont{R.}~\bibnamefont{Spekkens}}
\bibnamefont{and}
\bibinfo{author}{\bibfnamefont{T.}~\bibnamefont{Rudolph}},
\bibinfo{journal}{Phys. Rev. Lett.} \textbf{\bibinfo{volume}{89}},
\bibinfo{pages}{227901} (\bibinfo{year}{2002}).

\bibitem[{\citenamefont{Maassen and Uffink}(1988)}]{MU:genent}
\bibinfo{author}{\bibfnamefont{H.}~\bibnamefont{Maassen}}
\bibnamefont{and}
\bibinfo{author}{\bibfnamefont{J.}~\bibnamefont{Uffink}},
\bibinfo{journal}{Phys. Rev. Lett.} \textbf{\bibinfo{volume}{60}},
\bibinfo{pages}{1103} (\bibinfo{year}{1988}).

\bibitem[{\citenamefont{DiVincenzo
et~al.}(2001)\citenamefont{DiVincenzo, Leung, and
Terhal}}]{DLT:hiding}
\bibinfo{author}{\bibfnamefont{D.}~\bibnamefont{DiVincenzo}},
\bibinfo{author}{\bibfnamefont{D.}~\bibnamefont{Leung}},
\bibnamefont{and}
\bibinfo{author}{\bibfnamefont{B.}~\bibnamefont{Terhal}},
\bibinfo{journal}{IEEE Trans. Info. Theory}
\textbf{\bibinfo{volume}{48}}, \bibinfo{pages}{580}
(\bibinfo{year}{2001}), \bibinfo{note}{quant-ph/0103098}.

\bibitem[{foo()}]{footshort} \bibinfo{note}{Quantities like
$I(X;Y\!\tilde{Z}Z|Z\eqshort\tilde{Z})$ that appear in
Eq. (\ref{icrhoun}) and the following can be thought of as a shorthand
for $I(X;Y\!\tilde{Z}Z|w=1)$, where $W$ is a random variable whose
value is 1 if $Z=\tilde{Z}$ and is 0 otherwise.}

\bibitem[{\citenamefont{Wootters and Fields}(1989)}]{WF:mubs}
\bibinfo{author}{\bibfnamefont{W.}~\bibnamefont{Wootters}}
\bibnamefont{and}
\bibinfo{author}{\bibfnamefont{B.}~\bibnamefont{Fields}},
\bibinfo{journal}{Ann. Phys.} \textbf{\bibinfo{volume}{191}},
\bibinfo{pages}{368} (\bibinfo{year}{1989}).

\bibitem[{\citenamefont{Bandyopadhyay
et~al.}(2001)\citenamefont{Bandyopadhyay, Boykin, Roychowdhury, and
Vatan}}]{Bandyopadhyay01}
\bibinfo{author}{\bibfnamefont{S.}~\bibnamefont{Bandyopadhyay}},
\bibinfo{author}{\bibfnamefont{P.}~\bibnamefont{Boykin}},
\bibinfo{author}{\bibfnamefont{V.}~\bibnamefont{Roychowdhury}},
\bibnamefont{and}
\bibinfo{author}{\bibfnamefont{F.}~\bibnamefont{Vatan}}
(\bibinfo{year}{2001}), \bibinfo{note}{quant-ph/0103162}.

\bibitem[{\citenamefont{Sanchez}(1993)}]{sanchez}
\bibinfo{author}{\bibfnamefont{J.}~\bibnamefont{Sanchez}},
\bibinfo{journal}{Phys. Lett. A} \textbf{\bibinfo{volume}{173}},
\bibinfo{pages}{233} (\bibinfo{year}{1993}).

\bibitem[{\citenamefont{Fannes}(1973)}]{Fannes73a}
\bibinfo{author}{\bibfnamefont{M.}~\bibnamefont{Fannes}},
\bibinfo{journal}{Commun. Math. Phys.} \textbf{\bibinfo{volume}{31}},
\bibinfo{pages}{291} (\bibinfo{year}{1973}).

\bibitem[{\citenamefont{Ohya and Petz}(1993)}]{OhyaPetz}
\bibinfo{author}{\bibfnamefont{M.}~\bibnamefont{Ohya}}
\bibnamefont{and}
\bibinfo{author}{\bibfnamefont{D.}~\bibnamefont{Petz}},
\emph{\bibinfo{title}{Quantum Entropy and Its Use}}
(\bibinfo{publisher}{Springer-Verlag, Heidelberg},
\bibinfo{year}{1993}).

\bibitem[{\citenamefont{Schumacher and
Westmoreland}(2001)}]{Schumacher01}
\bibinfo{author}{\bibfnamefont{B.}~\bibnamefont{Schumacher}}
\bibnamefont{and} \bibinfo{author}{\bibfnamefont{M.~D.}
\bibnamefont{Westmoreland}} (\bibinfo{year}{2001}),
\bibinfo{note}{quant-ph/0112106}.

\end{thebibliography}


\pagebreak

\appendix
\section{Appendix}
\label{theappendix}
\subsection{Locking with more bases}

Intuitively, we expect a larger key to exert a stronger locking
effect (i.e., give a larger value of
$I_c^{(l)}(\rhol)-I_c(\rhol)$).
For instance, we have seen how classical mutual information can be
locked by encoding in one of two bases.  A natural question is,
can we lock more information by encoding in $L > 2$ bases?
A convenient choice of such bases are the mutually conjugate or
{\em mutually unbiased} bases, with the defining property that the
inner product between any two states from two different bases has
magnitude ${1 \over \sqrt{d}}$ in a $d$-dimensional system.
It is known that one can have at most $d+1$ mutually conjugate bases in
$d$ dimensions, and this maximum number of bases exists and can be
constructed when $d$ is a prime power~\cite{WF:mubs,
Bandyopadhyay01}. Let $U_1, \cdots, U_{d}$ take the computation
basis to each of these conjugate basis and $U_0 = I$.
In a scheme using $L$ bases (with key size $l = \log L$),
\bea
    \rhol = {1 \over Ld}
    \sum_{k=1}^d \sum_{t=0}^{L-1} (|k\>\<k| \ot |t\>\<t|)_A
    \ot (U_t |k\>\<k| U_t^\dag)_B
\,. \eea
When Alice tells Bob which basis $t$, the resulting state $\rhoun$
again has $I_c(\rhoun) = \log L + \log d = l + \log d$.
Applying the same analysis as before,
\bea
    \!\!\! I_c(\rhol) \leq \log d +
    \max_{|\phi\>} {1 \over L} \sum_{kt}
    |\<\phi|U_t|k\>|^2 \log| \<\phi|U_t|k\>|^2
\ss . \label{eq:iaccgenK} \eea
When $L=2$, $I_c(\rhol) = {1 \over L} \log d$.  Thus one would
hope $I_c(\rhol) = {1 \over L} \log d$ in general.
Unfortunately, the crucial entropic inequality in
Ref.~\cite{MU:genent} does not provide the desired bound.
Extensive numerical work on primes $3 \leq d \leq 29$ and $2 \leq L
\leq d+1$ shows that $I_c(\rhol) \approx ({1 \over L} + c) \log d$
where $c$ is roughly $0.1-0.15$ for the values of $d$ investigated.

In the extreme case of $L=d+1$, we can apply another entropic
inequality~\cite{sanchez} namely that the sum of the entropies is at
least $(d+1) \log \left({d+1\over 2}\right)$, so that
\bea
    & I_c(\rhol) \leq \log d - \log(d+1) + 1
    = 1-\log \lpm \!\! 1+{1\over d} \! \rpm & \mbox{and}
\non
\\
    & I_c(\rhoun) - I_c(\rhol) \geq 2 \log (d+1) - 1 \rule{0pt}{2.4ex}
    \hspace*{16ex}& \non
\eea
This still unlocks $\approx \log d$ bits, though the amount is
comparable to the $\log (d+1)$ bits communicated and thus we have no
(strong) violation of incremental proportionality in this regime.

\subsection{Small initial correlation}
\label{sec:boundedjump}

In Theorem 1, the difference between $I_c(\rho)$ and $I_c(\rho')$ is
bounded by the product of the initial correlation and the number of
different messages that can be sent.  Here, we bound the violation by
a function of the initial correlation only, allowing an arbitrary
number of qubits communicated interactively.  More formally, \\[1.3ex]
{\it Theorem 2} ~~
Let $\rho$ be a bipartite state on $C^d \otimes C^d$ and $\rho'$ be
obtained from $\rho$ by $l$ qubits of two-way communication.  Let
$d' < 2d$ be the least prime power no less than $d$, and 
$\eta(x) = -x \log x$.
If $I_c(\rho) \leq {1\over 6\ln 2}{1\over (d'+1)^2}$,
\bea
I_c(\rho') - I_c(\rho) \leq \hspace*{38ex} \non
\\
2l + 2 \ss (d'\!+\!\!1)^2 \!\! \sqrt{\ms 2 I_{\ms c \ms}(\ms\rho)
\ln \ms 2} \, \log \ms d \! + \eta \ss \bigl((d'\!+\!\!1)^2 \!\!
\sqrt{2 I_{\ms c \ms}(\ms \rho) \ln 2} \bigr) \ss. \!\!\!\!
\non
\eea

\noindent A simpler, but less tight expression can be obtained from
the above by expanding the $\log$ function in $\eta$:
\bea I_c(\rho') -  I_c(\rho) \hspace*{38ex}
\non
\\
\leq 2l - (d'\!+\!\!1)^2 \!\!
\sqrt{\ms (2 \ln 2) I_{\ms c \ms}(\ms\rho) } \,
\log \! \sqrt{\ms (2 \ln 2) I_{\ms c \ms}(\ms\rho)} \, .
\non
\eea

Thus even for the most general communication model, incremental
proportionality violation is continuous in the initial correlations,
with incremental proportionality holding when the initial state is
uncorrelated (a special case of pure initial states).  However the
present bound is not uniform with respect to the size of the support
of $\rhol$: To get $I_c(\rhoun)\leq 2l+\delta$, we need approximately
$I_c(\rhol)\leq \delta^2 (2d)^{-4}$.

\vspace*{1ex}

\noindent {\em Proof:} The theorem can be proved by putting
together various properties of $I_c(\rho)$ and $I_q(\rho)$, and
the main steps of the proof can be summarized as:
\bea
    \hspace*{-2ex} & & \hspace*{-0.5ex} I_c(\rho') - I_c(\rho)
    \leq I_c(\rho')
    \stackrel{1}{\leq} I_q(\rho')
    \stackrel{2}{\leq} 2l + I_q(\rho)
\non
\\
    \hspace*{-2ex} & \stackrel{3}{\leq} & \hspace*{-0.5ex}
    2l + \log \ms d^2 \,
    \tr|\rho_{A} \! \ot \! \rho_{B} \! - \! \rho_{AB}| +
    \eta(\tr|\rho_{A} \! \ot \! \rho_{B} \! - \! \rho_{AB}|)
\non
\\
    \hspace*{-2ex} & \stackrel{4}{\leq} & \hspace*{-0.5ex}
    2l \ms + \ms 2 (\log \ms d) (d'\!\!+\!\!1)^2 \!\ms
        \sqrt{\ms 2 I_{\ms c \ms}(\ms\rho\ms) \ln \ms 2} +
    \eta \ss ((d'\!\!+\!\!1)^2 \!\!
    \sqrt{\ms 2 I_{\ms c \ms}(\ms\rho\ms) \ln \ms 2})
\non \eea

First, we explain the intuition behind the properties that make
each step valid, then, we complete the proof by proving each of
the steps.
The idea is to upper bound $I_c$ by $I_q$ and use incremental
proportionality of $I_q$ \cite{terhal+:epur} in steps 1 and 2.  Then
it remains to show that $I_q$ is small if $I_c$ is small, and this is
done in steps 3 and 4.  Step 3 expresses $I_q$ as a difference of the
entropies $S(\rho)$ and $S(\rho_A \ot \rho_B)$, which is subsequently
bounded by Fannes' inequality \cite{Fannes73a}.  Step 4 is to prove
and apply the following lemma:
\begin{lem}
\label{lem:tr-Ic} If $\rho$ is a bipartite state on $C^{d} \otimes
C^d$, then \bea \hspace*{2ex} \tr \ss | \ss \rho_{AB} - \rho_A \ot
\rho_B|
    \leq (d'+1)^2 \sqrt{2 \ln 2 ~ I_c(\rho)}
\label{eq:trace-Ic} \,, \eea where $d' < 2d$ is a prime power
no less than $d$.
\end{lem}
This lemma says that a state with small classical mutually
information is close to being a product state, and a simple
consequence is that $I_c(\rho) = 0$ iff $\rho$ is a product state.
Steps 3 and 4 give the desired bound of $I_q$ in terms of $I_c
\,$: If $I_c \leq {1\over 6\ln 2}{1\over (d'+1)^2}$ then
\bea I_q \leq 2 \ss (d'\!+\!1)^2 \! \! \sqrt{2 I_{\ms c
\ms}(\ms\rho\ms) \ln 2} \, \log d \! + \eta((d'\!+\!1)^2 \! \!
\sqrt{2 I_{\ms c \ms}(\ms\rho\ms) \ln 2})
\non \,. \eea

\noindent We proceed to prove steps 1, 3, and 4.  First, $I_c$ and
$I_q$ can be rewritten as \cite{OhyaPetz}:
\bea
    I_q(\rho) & = & S(\rho_{AB} \| \ss \rho_A \ot \rho_B)
\label{eq:facta} \,,
\\[1.2ex]
    I_c(\rho) & = & \max_{M_A \ot M_B}
    S(p_{AB} \| \ss p_A \ot p_B)
\label{eq:factc} \,, \eea
where $p_{AB}, p_A, p_B$ are the probability distributions of the
joint and individual outcomes of applying a local measurement $M_A
\otimes M_B$ to $\rho$, and the quantum relative entropy is
defined as
\be
    S(\nu \| \ss \mu) := \tr(\nu \log \nu) - \tr(\nu \log \mu)
\,. \ee

To prove step 1, let $M_A$ and $M_B$ be the optimal measurements
for $I_c(\rho)$.
Let $\Lambda$ be the local quantum operation of applying $M_A \ot
M_B$ followed by storing the classical outcomes in ancillas $A'$
and $B'$ and discarding of the original systems $A$ and $B$.
The final state $\tilde{\rho}_{A'B'} = \Lambda(\rho)$ is a
classical state corresponding to $p_{AB}$ so that
\bea
    I_c(\rho) & = & S(p_{AB} \| \ss p_A \ot p_B)
\non
\\
    & = & S(\tilde{\rho}_{A'B'} \| \ss
    \tilde{\rho}_{A'} \ot \tilde{\rho}_{B'})
\non
\\
    & \leq & S(\rho_{AB} \| \ss \rho_{A} \ot \rho_{B})
    \; = \; I_q(\rho) \,, 
\label{eq:sadd} \eea
where the inequality in \eq{sadd} is due to monotonicity of
$I_q(\rho)$ under the local operation $\Lambda$.

To prove step 3, recall Fannes' inequality~\cite{Fannes73a} for
$d_0$-dimensional states $\nu$, $\mu$ with $\tr|\nu-\mu| \leq
1/e$,
\bea
    |S(\nu) \! - \! S(\mu)| \leq \log d_0 \ss
    \tr|\nu-\mu| + \eta(\tr|\nu-\mu|) \hspace*{-4ex}
\non \eea
Hence, if $\tr|\rho_{A} \! \ot \! \rho_{B} - \rho_{AB}|\leq 1/e$,
\bea
    I_q(\rho) = S(\rho_{A} \! \ot \! \rho_{B}) \! - \! S(\rho_{AB})
    \hspace*{23ex}
\non
\\
    \leq 2 \log \ms d \,
    \tr|\rho_{A} \! \ot \! \rho_{B} \! - \! \rho_{AB}| +
    \eta(\tr|\rho_{A} \! \ot \! \rho_{B}\! - \!\rho_{AB}|) \ss.~
\label{eq:fromfannes} \eea

Once Lemma \ref{lem:tr-Ic} is proved, step 4 can be obtained by
substituting \eq{trace-Ic} into \eq{fromfannes}.
We prove Lemma \ref{lem:tr-Ic} for $\rho$ on $C^d \otimes C^d$
where $d = p^n$ is a prime power.  The general case follows,
because when $d$ is not a prime power, $\rho$ can still be taken
as a state on $C^{d'} \otimes C^{d'}$ where $d' < 2d$ is the
least prime power no less than $d$.
The main idea in the proof is to rewrite $\tr|\rho_{AB}\!-\!\rho_A
\ot \rho_B|$ as a sum, each term of which is bounded by the
initial classical mutual information of $\rho$.
This relies on the following result proved for
$d=p^n$~\cite{WF:mubs,Bandyopadhyay01}.
There exists a basis $\{M_k^i\}_{k=1,\cdots,d+1, i=1,\cdots,d-1}$
for traceless $d \times d$ matrices, such that $\tr M_k^{i
\dagger} M_l^j = d \delta_{kl} \delta_{ij}$, i.e. orthonormal
under the trace norm up to a scaling factor.  Furthermore, for
each $k$, $\{M_k^i\}_{i=1,\cdots,d-1}$ is a commuting set, and can
be simultaneously diagonalized by conjugation by some
$U_k^\dagger$.

Using $\{M^k_i\}_{i,k} \cup \{I\}$ as a basis for $d \times d$
matrices, we can express $\rho_{AB}$ as
\bea
    \rho_{AB} & = & {1 \over d^2} \lbL
    \!\! I \! \ot \! I + \sum_{ki} \alpha_{ki00} \, M^k_i \! \ot \! I
    + \sum_{lj} \alpha_{00lj} \, I \! \ot \! M^l_j
\non
\\     & & \hspace*{5ex} + \sum_{kilj} \alpha_{kilj} \, M^k_i \ot M^l_j \rbL
\eea
with generally complex coefficients $\alpha_{kilj}$.
Using the commutivity of each $\{M^k_i\}_i$,
\bea
    \rho_{AB} & \!= \!& {1 \over d^2} \lbL \!\! I \! \ot \! I +
    \sum_k (U_k E_k U_k^\dag) \! \ot \! I +
    \sum_l I \! \ot \! (U_l F_l U_l^\dag)
\non
\\
    & & \hspace*{5ex} + \sum_{kl} (U_k \! \ot \! U_l)
    D_{kl} (U_k \! \ot \! U_l)^\dagger \! \rbL
\eea
for some diagonal matrices $E_k$, $F_k$, and $D_{kl}$ where each
$E_k$, $F_k$ is $d \times d$, and each $D_{kl}$ is $d^2 \times
d^2$.
Then,
\bea
    \rho_A \!=\! {1 \over d} \lbL I \!+\! \sum_k U_k E_k U_k^\dagger \rbL
\,,~
    \rho_B \!=\! {1 \over d} \lbL I \!+\! \sum_k U_k F_k U_k^\dagger \rbL
\,, \non \eea
and
\bea
    \rho_{AB} \! - \! \rho_A \ot \rho_B
    \! =\!  {1 \over d^2} \!
    \sum_{kl} (U_k \! \ot \! U_l)
    (D_{kl} \! - \! E_k \! \ot \! F_k) (U_k \! \ot \! U_l)^\dagger
\non
\\
    \tr \, | \, \rho_{AB} - \rho_A \ot \rho_B | \leq
    {1 \over d^2} \sum_{kl} \tr \, | \, D_{kl} - E_k \ot F_k | \hspace*{8ex}
\non
\\ [-0.3ex]
    ~ = ~ \sum_{kl} \tr |\, p_{kl} - q_{kl} |
    ~ \leq ~ (d+1)^2 \, \sqrt{2 \ln 2 ~ S(p_{kl}||q_{kl})}  \,,
\non
\eea
where the last inequality follows from $S(\nu \| \ss \mu) \geq {1
\over 2 \ln 2} (\tr|\nu - \mu|)^2$ \cite{OhyaPetz,Schumacher01}.
In the above, $p_{kl}$ and $q_{kl}$ are probability distributions
of the outcomes when locally measuring $\rho$ and $\rho_A \ot
\rho_B$ along the simultaneously tensor product eigenbasis of
$\{M_i^k\}_i$ and $\{M_i^l\}_i$.  In each of these measurements,
only the $kl$ terms contribute due to the orthonormality of the
basis chosen.
According to \eq{factc}, the relative entropy between $p_{kl}$ and
$q_{kl}$ is a lower bound for $I_c$.  Thus,
\be \tr \, | \, \rho_{AB} - \rho_A \ot \rho_B | \leq (d+1)^2 \,
\sqrt{2 \ln 2 ~ I_c(\rho)} \ee
This completes the proofs for all the steps and thus our theorem.

\end{document}